\documentclass[11pt]{article}
\usepackage{amsmath}
\usepackage{fullpage}
\usepackage{hyperref}
\usepackage{amsfonts}
\usepackage{amssymb}
\newtheorem{theorem}{Theorem}

\newtheorem{lemma}[theorem]{Lemma}

\newtheorem{remark}[theorem]{Remark}

\newenvironment{proof}[1][Proof]{\par\noindent{\textsl{#1.} }}%
{\hfill$\rule{0.5em}{0.5em}$ \par\bigbreak}

\newcommand{\R}{\ensuremath{\mathbb{R}}}

\newcommand{\F}{\ensuremath{\mathbb{F}}}

\renewcommand{\a}{\alpha}
\newcommand{\bp}{\mathsf{MP}}
\newcommand{\ldpc}{\mathsf{LDPC}}
\newcommand{\bec}{\mathsf{BEC}}
\newcommand{\bsc}{\mathsf{BSC}}
\newcommand{\biawgn}{\mathsf{BIAWGN}}
\newcommand{\eps}{\varepsilon}
\renewcommand{\epsilon}{\varepsilon}

\title{Iterative Decoding of Low-Density Parity Check Codes\thanks{This is a survey written for the Computational Complexity Column of the Bulletin of the European Association for Theoretical Computer Science (EATCS), Issue 90, October 2006.} \\
  {\normalsize (An
  Introductory Survey)}}

\author{Venkatesan Guruswami\thanks{Supported in part by NSF CCF-0343672, an Alfred P. Sloan Research Fellowship, and a David and Lucile Packard Foundation
    Fellowship.}}

\date{Department of Computer Science and
    Engineering \\ University of Washington \\ Seattle, WA 98195 ~ \\ ~ \\ September 2006}

\begin{document}

\maketitle

\begin{abstract}\noindent
Much progress has been made on decoding algorithms for
error-correcting codes in the last decade. In this article, we give an
introduction to some fundamental results on iterative, message-passing
algorithms for low-density parity check codes. For certain
important stochastic channels, this line of work has enabled getting
very close to Shannon capacity with algorithms that are extremely
efficient (both in theory and practice).
\end{abstract}

\newpage
\tableofcontents

\newpage
\section{Introduction}

Over the past decade or so, there has been substantial new progress on
algorithmic aspects of coding theory. A (far from exhaustive) list of
the themes that have witnessed intense research activity includes:
\begin{enumerate}
\item A resurgence of interest in the long forgotten class of
low-density parity check (LDPC) codes and on iterative,
message-passing decoding algorithms for them, which has resulted in
codes with rates extremely close to Shannon capacity together with
efficient decoding algorithms.
\item Linear time encodable/decodable error-correcting codes (based on expanders) for worst-case errors.
\item List decoding algorithms which correct many more worst-case
errors beyond the ``half-the-code-distance'' bound, and which can
achieve capacity even against adversarial noise.\footnote{The
capacity-achieving part was recently shown for codes over {\em large}
alphabets, specifically explicit codes of rate close to $1-p$ that can
be list decoded in polynomial time from a fraction $p$ of errors were
constructed in \cite{GR-capacity}.  For binary codes, the capacity for
decoding a fraction $p$ of errors equals $1-H(p)$, but we do not know
how to achieve this constructively.}
\end{enumerate}
Of course there are some interrelations between the above directions;
in particular, progress on linear-time encodable/decodable codes is
based on expander codes, which are LDPC codes with additional
properties. Also, list decoding algorithms that run in linear time and
correct a fraction $\rho$ of errors for any desired $\rho < 1$ have
been developed using expander-based ideas~\cite{GI-linld}.  

Of the above lines of work, the last two have a broader following in
the theoretical computer science community, due to their focus on the
combinatorial, worst-case noise model and the extraneous applications
of such codes in contexts besides communication (such as
pseudorandomness and average-case complexity). The sister complexity
theory column that appears in SIGACT news featured recent surveys on
both these topics~\cite{G-sigactsurvey,sudan-sigact}. A longer survey
on very recent developments in list decoding of algebraic codes will
appear in \cite{Gur-nowsurvey}. A very brief survey featuring couple
of complexity-theoretic uses of list decoding appears in
\cite{Gur-ITW}.  Applications of coding theory to complexity theory,
especially those revolving around sub-linear algorithms, are surveyed
in detail in \cite{luca-survey}.

We use the opportunity provided by this column to focus on the first
line of work on iterative (also called message-passing or belief
propagation) algorithms for decoding LDPC codes. This is in itself a
vast area with numerous technically sophisticated results.  For a
comprehensive discussion of this area, we point the reader to the
upcoming book by Richardson and Urbanke~\cite{RU-book}, which is an
excellent resource on this topic.  The February 2001 issue of Volume
47 of the IEEE Transactions on Information Theory is another valuable
resource --- this was a special issue dedicated to iterative decoding
and in particular contains the series of papers
\cite{LMSS,LMSS-errors,RU01,RSU01}. This sequence of papers is
arguably one of the most important post-Gallager developments in the
analysis of iterative decoding, and it laid down the foundations for
much of the recent progress in this field.

\smallskip \noindent {\bf Disclaimer:} The literature on the subject
of LDPC and related codes and belief propagation algorithms is vast
and diverse, and the author, not having worked on the topic himself,
is only aware of a small portion of it. Our aim will be to merely
provide a peek into some of the basic context, results, and methods of
the area.  We will focus almost exclusively on LDPC codes, and
important related constructions such as LT codes, Raptor codes,
Repeat-Accumulate codes, and turbo codes are either skipped or only
very briefly mentioned.  While the article should (hopefully) be
devoid of major technical inaccuracies, we apologize for any
inappropriate omissions in credits and citations (and welcome comments
from the reader if any such major omissions are spotted).

\medskip \noindent {\bf Organization:}
We begin with some basic background information concerning LDPC
codes, the channel models we will study, and the goal of this line of
study in Section~\ref{sec:backgd}. In Section~\ref{sec:concat}, we
discuss how concatenated codes with an outer code that can correct a
small fraction of errors can be used to approach capacity, albeit with
a poor dependence on the gap to capacity. We then turn to message
passing algorithms for LDPC codes and describe their high level
structure in Section~\ref{sec:MP-highlevel}. With this in place, we develop and
analyze some specific message passing algorithms for {\em regular}
LDPC codes in Section~\ref{sec:regular}, establishing theoretical
thresholds for the binary erasure and binary symmetric channels. We then turn our focus to {\em
  irregular} LDPC codes in Section~\ref{sec:irregular}, and discuss,
among other things, how one can use them to achieve the capacity of
the binary erasure channel. Finally, in Section~\ref{sec:ira}, we discuss how one can achieve linear
encoding time for LDPC codes, and also discuss a variant called
Irregular Repeat-Accumulate (IRA) codes that are linear-time encodable
by design and additionally offer improved complexity-vs-performance trade-offs.

\section{Background}
\label{sec:backgd}
\subsection{Linear and LDPC codes}
We will focus exclusively on binary linear codes. A binary linear code
$C$ of {\em block length} $n$ is a subspace of $\F_2^n$ where $\F_2 =
\{0,1\}$ is the field with two elements. The rate of $C$, denoted
$R(C)$, equals $k/n$ where $k$ is the dimension of $C$ (as a vector
space over $\F_2$); such a code is also referred to as an $[n,k]$
code. Being a linear subspace of dimension $k$, the code $C$ can be
described as the kernel of a matrix $H \in \F_2^{(n-k) \times n}$, so
that $C = \{ c \in \F_2^n \mid Hc = 0\}$ (we treat codewords $c$ as
column vectors for this description). The matrix $H$ is called the
{\em parity check matrix} of the code $C$. In general, any choice of
$H$ whose rows form a basis of the dual space $C^{\perp} = \{ x \in
\F_2^n \mid x^t c = 0 \forall c \in C\}$ describes the same code. Of
special interest to us here are codes that admit a {\em sparse} parity
check matrix. In particular, we will study {\em low-density parity
check} (LDPC) codes, which were introduced and studied in Gallager's
amazing work~\cite{gallager} that was way ahead of its time. LDPC
codes are described by a parity check matrix all of whose rows and
columns have at most a fixed constant number of $1$'s (the constant is
independent of $n$).\footnote{We will throughout be interested in a
family of codes of increasing block length $n$ with rate $k/n$ held a
fixed constant. For convenience, we don't spell this out explicitly,
but this asymptotic focus should always be kept in mind.}

A convenient way to describe an LDPC code is in terms of its {\em
  factor graph}.\footnote{This graphical representation applies for
  any linear code. But the resulting graph will be sparse, and hence
  amenable to linear time algorithms, only for LDPC codes.} This is a
  natural bipartite graph defined as follows. On the left side are $n$
  vertices, called {\em variable} nodes, one for each codeword
  position. On the right are $m = n-k$ vertices, called {\em check}
  nodes, one for each parity check (row of the parity check matrix). A
  check node is adjacent to all variable nodes whose corresponding
  codeword symbols appear in this parity check. In other words, the
  parity check matrix of the code is precisely the bipartite adjacency
  matrix of the factor graph.
  
  A special class of LDPC codes are {\rm regular} LDPC codes where the
  factor graph is both left-regular and right-regular. Regular LDPC
  codes were in fact the variant originally studied by
  Gallager~\cite{gallager}, as well as in the works of Mackay and
  Neal~\cite{mackay,MN} and Sipser and Spielman~\cite{SS96,spielman96}
  that sparked the resurgence of interest in LDPC codes after over 30
  years since Gallager's work.\footnote{In the long interim period,
    LDPC codes went into oblivion, with the exception of two (known to
    us) works. Zyablov and Pinsker~\cite{ZP-ldpc} proved that for
    random LDPC codes, with high probability over the choice of the
    code, Gallager's algorithm corrected a constant fraction of {\em
      worst-case} errors. Tanner~\cite{tanner} presented an important
    generalization of Gallager's construction and his decoding
    algorithms, which was later important in the work on linear time
    decodable expander codes~\cite{SS96}.}
LDPC codes based on non-regular graphs, called
 irregular LDPC codes, rose to prominence beginning in the work of
 Luby {\it et al}~\cite{LMSS,LMSS-errors} (studying codes
 based on irregular graphs was one of the big conceptual leaps made in
 these works). We will return to this aspect later in the survey. A
 popular choice of regular LDPC codes (with a rate of $1/2$) are
 $(3,6)$-regular LDPC codes where variable nodes have degree $3$ and
 check nodes have degree $6$.

\subsection{Channel models and their capacity}
Design of good LDPC codes, together with progress in analyzing natural
message-passing algorithms for decoding them, has led to rapid
progress towards approaching the capacity of important stochastic
channels. We now review the main noise models that we will be
interested in.

Throughout, we deal with binary codes only. We will
find it convenient to use $\{+1,-1\}$ (instead of $\{0,1\}$) for the
binary alphabet, where $+1$ corresponds to the bit $0$ and $-1$ to the
bit $1$. Note the XOR operation becomes multiplication in the $\pm 1$
notation.

We will assume the channel's operation to be {\em memoryless}, so that
each symbol of the codeword is distorted independently
according to the same channel law. So to specify the noise model, it
suffices to specify how the noise distorts a single input symbol. For
us the input symbol will always be either $\pm 1$, and so the channels
have as input alphabet ${\cal X} = \{1,-1\}$. Their output alphabet
will be denoted by ${\cal Y}$ and will be different for the different
channels. Upon transmission of a codeword $c \in {\cal X}^n$, the word
$y$ observed by the receiver belongs to ${\cal Y}^n$. The receiver
must then decode $y$ and hopefully compute the original transmitted
codeword $c$. The challenge is to achieve a vanishingly small error
probability (i.e., the probability of either a decoding failure or an
incorrect decoding), while at the same time operating at a good rate,
hopefully close to the capacity of the channel.

We begin with the simplest noise model, the {\em Binary Erasure
Channel} (BEC). This is parameterized by a real number $\a$, $0 \le \a
< 1$. The output alphabet is ${\cal Y} = \{1,-1,?\}$, with $?$
signifying an {\em erasure}. Upon input $x \in {\cal X}$, the channel
outputs $x$ with probability $1-\a$, and outputs $?$ with probability
$\a$. The value $\a$ is called the erasure probability, and we denote
by $\bec_\a$ the BEC with erasure probability $\a$. For large $n$,
the received word consists of about $(1-\a) n$ unerased symbols with
high probability, so the maximum rate at which reliable communication
is possible is at most $(1-\a)$ (this holds even if the transmitter
and receiver knew in advance which bits will be erased). It turns out
this upper bound can be achieved, and Elias~\cite{elias-erasure}, who
first introduced the BEC, also proved that its capacity equals
$(1-\a)$.

The {\em Binary Symmetric Channel} (BSC) is parameterized by a real
number $p$, $0 \le p < 1/2$, and has output alphabet ${\cal Y} =
\{1,-1\}$. On input $x \in {\cal X}$, the channel outputs $bx$ where
$b = -1$ with probability $p$ and $b=1$ with probability $1-p$. The
value $p$ is called the {\em crossover probability}. The BSC with
crossover probability $p$ is denoted by $\bsc_p$. The capacity of
$\bsc_p$ is well known to be $1- H(p)$, where $H(p) = - p \lg p -
(1-p) \lg (1-p)$ is the binary entropy function.

Finally, we mention a channel with continuous output alphabet ${\cal
  Y}$ called {\em Binary Input Additive White Gaussian Noise}
  (BIAWGN). Here ${\cal Y}$ equals the set of real numbers, and the
  channel operation is modeled as $y = x+z$ where $x \in \{\pm1\}$ is
  the input and $z$ is a normal variable with mean $0$ and variance
  $\sigma^2$ (i.e., has probability density function $p(z) =
  \frac{1}{\sqrt{2\pi\sigma^2}} e^{-\frac{z^2}{2\sigma^2}}$). We denote
  by $\biawgn_{\sigma}$ the BIAWGN with variance $\sigma^2$; its
  capacity is a function of $1/\sigma^2$ alone, though there is no
  elementary form expression known for the capacity (but it can be
  expressed as an integral that can be estimated numerically). For
  rate $1/2$, the largest $\sigma$ (Shannon limit) for which reliable
  communication on the BIAWGN channel is possible is (up to the
  precision given) $\sigma_{\rm opt} = 0.9787$. 

More generally, if we allow scaling of inputs, the capacity is a
  function of the ``signal-to-noise'' ratio $E_N/\sigma^2$ where $E_N$
  is the energy expended per channel use. If the inputs to the channel
  are not constrained to be $\pm 1$, but instead can take arbitrary
  real values, then it is well known that the capacity of the AWGN
  channel equals $\frac{1}{2} \log_2 \left(1 + E_N/\sigma^2 \right)$
  bits per channel use. In particular, in order to achieve reliable
  communication at a rate of $1/2$ over the real-input AWGN channel, a
  signal-to-noise ratio of $1$, or $0$ dB, is required.\footnote{In
  decibel notation, $\lambda > 0$ is equivalent to $10 \log_{10}
  \lambda$ dB.}  For the BIAWGN channel, this ratio increases to
  $1/\sigma_{\rm opt}^2 = 1.044$ or $0.187$ dB. Accordingly, the
  yardstick to measure the quality of a decoding algorithm for an LDPC
  code of rate $1/2$ is how close to this limit it can lead to correct
  decoding with probability tending to $1$ (over the realization of
  the BIAWGN channel noise).

The continuous output of a BIAWGN channel can be quantized to yield a
discrete approximation to the original value, which can then be used
in decoding. (Of course, this leads to loss in information, but is
often done for considerations of decoding complexity.) A particularly
simple quantization is to decode a signal $x$ into $1$ if $x \ge 0$
and into $-1$ if $x < 0$. This effectively converts an AWGN channel
with variance $\sigma^2$ into a BSC with crossover probability
$Q(1/\sigma) = \frac{1}{\sqrt{2\pi}} \int_{1/\sigma}^{\infty}
e^{-x^2/2} dx$. It should not come as a surprise that the capacity of
the resulting BSC falls well short of the capacity of the BIAWGN.

All the above channels have the following {\em output-symmetry}
property: For each possible channel output $q$, $p(y=q | x=1) = p(y =
-q|x=-1)$. (Here $p(y|x)$ denotes the conditional probability that the
channel output equals $y$ given the channel input is $x$.)

We will focus a good deal of attention on the BEC. Being a very simple
channel, it serves as a good warm-up to develop the central ideas, and at
the same time achieving capacity on the BEC with iterative decoding of
LDPC codes is technically non-trivial. The ideas which were originally
developed for erasure codes in \cite{LMSS} have been generalized for
more general channels, including the BSC and BIAWGN, with great
success~\cite{LMSS-errors,RU01,RSU01}. Yet, to date the BEC is the
only channel known for which one can provably get arbitrarily close to
capacity via iterative decoding of (an ensemble of) LDPC
codes. So naturally, given our focus on the theoretical aspects, the
BEC is of particular interest.

\subsection{Spirit of the results}

The central goal of research in channel coding is the following: given
a particular channel, find a family of codes which have fast (ideally
linear-time) encoding algorithms and which can be reliably decoded in
linear time at rates arbitrarily close to channel capacity. This is, of
course, also the goal of the line of work on LDPC codes.

In ``practice'' one of the things that seems to get people excited are
plots of the signal-to-noise ratio (SNR) vs bit error probability
(BER) for finite-length codes found by non-trivial optimization based
on theoretical insights, followed by simulation on, say, the BIAWGN
channel. Inspired by the remarkable success on the BEC~\cite{LMSS},
this approach was pioneered for LDPC codes in the presence of errors in
\cite{spielman-allerton,LMSS-errors}, culminating in the demonstration
of codes for the BIAWGN channel in \cite{RSU01} that beat turbo codes
and get very close to the Shannon limit.  

Since this article is intended for a theory audience, our focus will
be on the ``worst'' channel parameter (which we call threshold) for
which one can prove that the decoding will be successful with
probability approaching $1$ in the asymptotic limit as the block
length grows to infinity.  The relevant channel parameters for the
BEC, BSC, and BIAWGN are, respectively, the erasure probability,
crossover probability, and the variance of the Gaussian noise.  The
threshold is like the random capacity for a {\em given} code (or
ensemble of codes) and a {\em particular} decoder. Normally for
studying capacity we fix the channel and ask what is the largest rate
under which reliable communication is possible, whereas here we fix
the rate and ask for the worst channel under which probability of
miscommunication tends to zero.  Of course, the goal is to attain as a
large a threshold as possible, ideally approaching the Shannon limit
(for example, $1-\a$ for $\bec_\alpha$ and $1-H(p)$ for $\bsc_p$).

\iffalse
While the
analytic bounds themselves may not be directly practically appealing
due to large constants and block length, they do lead to the correct
insights for designing good codes, and the resulting constructions
perform better in practice than predicted by the analytic bounds.  The
underlying algorithms will always be simple and practical iterative
methods; the challenge is in the analysis. The goal is to attain as
large a threshold as possible, and ideally approaching capacity. The
ultimate problem: linear-time encoding, linear-time decoding at rates
arbitrarily close to capacity. A theoretical formulation will be to
have runtime linear in $n$ and polynomial in $1/\eps$, the distance to
capacity.}
\fi

\section{Simple concatenated schemes to achieve capacity on BEC and BSC}
\label{sec:concat}
We could consider the channel coding problem solved (at least in
theory) on a given channel if we have explicit codes, with efficient
algorithms for encoding and reliable decoding at rates within any
desired $\eps$ of capacity. Ideally, the run time of the algorithms
should be linear in the block length $n$, and also depend polynomially
on $1/\eps$. (But as we will see later, for certain channels like the
BEC, we can have a runtime of $O(n \log(1/\eps))$, or even better $c
n$ with $c$ independent of $\eps$, if we allow randomization in the
construction.) In this section, we discuss some ``simple'' attacks on
this problem for the BEC and BSC, why they are not satisfactory, and
the basic challenges this raises (some of which are addressed by the
line of work on LDPC codes).

For the BEC, once we have the description of the generator matrix of a
linear code that achieves capacity, we can decode in $O(n^3)$ time by
solving a linear system (the decoding succeeds if the system has a
unique solution). Since a random linear code achieves capacity with
high probability~\cite{elias-erasure}, we can sample a random
generator matrix, thus getting a code that works with high probability
(together with a cubic time algorithm). However, we do not know any
method to {\em certify} that the chosen code indeed achieves capacity.
The drawbacks with this solution are the cubic time and
randomized nature of the construction.

A construction using {\em concatenated codes} gets around both these
shortcomings. The idea originates in Forney's
work~\cite{forney-thesis} that was the first to present codes
approaching capacity with polynomial time encoding and decoding
algorithms. 

Let $\a$ be the erasure probability of the BEC and say our goal is to
construct a code of rate $(1-\a-\eps)$ that enables reliable
communication on $\bec_\a$.  Let $C_1$ be a linear time
encodable/decodable binary code of rate $(1-\eps/2)$ that can correct
a small constant fraction $\gamma = \gamma(\eps) > 0$ of {\em
worst-case} erasures. Such codes were constructed in
\cite{spielman96,AL-erasures}. For the concatenated coding, we do the
following. For some parameter $b$, we block the codeword of $C_1$ into
blocks of size $b$, and then encode each of these blocks by a suitable
{\em inner} binary linear code $C_2$ of dimension $b$ and rate
$(1-\a-\eps/2)$. The inner code will be picked so that it achieves the
capacity of the $\bec_\a$, and specifically recovers the correct
message with success probability at least $1-\gamma/2$. For $b =
b(\eps,\gamma) = \Omega\left(\frac{\log(1/\gamma)}{\eps^2}\right)$, a
random code meets this goal with high probability, so we can find one
by brute-force search (that takes constant time depending only on
$\eps$).

The decoding proceeds as one would expect: first each of the inner
blocks is decoded, by solving a linear system, returning either
decoding failure or the correct value of the block. (There are no
errors, so when successful, the decoder knows it is correct.) Since
the inner blocks are chosen to be large enough, each inner decoding
fails with probability at most $\gamma/2$. Since the noise on
different blocks are independent, by a Chernoff bound, except with
exponentially small probability, we have at most a fraction $\gamma$
of erasures in the outer codeword. These are then handled by the
linear-time erasure decoder for $C_1$.

We conclude that, for the $\bec_\a$, we can construct codes of rate
$1-\a-\eps$, i.e., within $\eps$ of capacity, that can be encoded and
decoded in $n/\eps^{O(1)}$ time. While this is pretty good, the
brute-force search for the inner code is unsatisfying, and the BEC is
simple enough that better runtimes (such as $O(n \log(1/\eps))$) are
achieved by certain irregular LDPC codes.

A similar approach can be used for the $\bsc_p$. The outer code $C_1$
must be picked so that it can correct a small fraction of worst-case
{\em errors} --- again, such codes of rate close to $1$ with linear
time encoding and decoding are known~\cite{spielman96,GI-zyablov}.
Everything works as above, except that the decoding of the inner
codes, where we find the codeword of $C_2$ closest to the received
block, requires a brute-force search and this takes
$2^b = 2^{\Omega(1/\eps^2)}$ time. This can be improved to polynomial in
$1/\eps$ by building a look-up table, but then the size of the look-up
table, and hence the space complexity and time for precomputing the
table, is exponential in $1/\eps$.

In summary, for the $\bsc_p$, we can construct codes of rate
$1-H(p)-\eps$, i.e., within $\eps$ of capacity, that can be encoded in
$n/\eps^{O(1)}$ time and which can be reliably decoded in $n
2^{1/\eps^{O(1)}}$ time. It remains an important open question to
obtain such a result with decoding complexity $n/\eps^{O(1)}$, or even
${\rm poly}(n/\eps)$.\footnote{We remark that asymptotically, with
  $\eps$ fixed and $n \to \infty$, the exponential dependence on
  $1/\eps$ can be absorbed into an additional factor with a slowly
  growing dependence on $n$. However, since in practice one is
  interested in moderate block length codes, say $n \le 10^6$, a
  target runtime such as $O(n/\eps)$ seems like a clean way to pose
  the underlying theoretical question.}

We also want to point out that recently an alternate method using LP
decoding has been used to obtain polynomial time decoding at rates
arbitrarily close to capacity~\cite{FS-lpdecoding}. But this also suffers
from a similar poor dependence on the gap $\eps$ to capacity.

\section{Message-passing iterative decoding: An abstract view}
\label{sec:MP-highlevel}
\subsection{Basic Structure}
We now discuss the general structure of natural message-passing
iterative decoding algorithms, as discussed, for example, in
\cite{RU01}. In these algorithms, messages are exchanged between the
variable and check nodes in discrete time steps.  Initially, each
variable node $v_j$, $1 \le j \le n$, has an associated received value
$r_j$, which is a random variable taking values in the channel output
alphabet ${\cal Y}$. Based on this, each variable sends a message
belong to some message alphabet ${\cal M}$. A common choice for this
initial message is simply the received value $r_j$, or perhaps some
quantized version of $r_j$ for continuous output channels such as
BIAWGN. Now, each check node $c$ processes the messages it receives
from its neighbors, and sends back a suitable message in ${\cal M}$ to
each of its neighboring variable nodes. Upon receipt of the messages
from the check nodes, each variable node $v_j$ uses these together
with its own received value $r_j$ to produce new messages that are
sent to its neighboring check nodes. This process continues for many
time steps, till a certain cap on the number of iterations is
reached. In the analysis, we are interested in the probability of
incorrect decoding, such as the bit-error probability.  For every time
step $i$, $i \in {\mathbb N}$, the $i$'th iteration consists of a
round check-to-variable node messages, followed by the variable nodes
responding with their messages to the check nodes. The $0$'th
iteration consists of dummy messages from the check nodes, followed by
the variable nodes sending their received values to the check nodes.

A very important condition in the determination of the next message
based on the messages received from the neighbors is that message sent
by $u$ along an edge $e$ {\em does not depend on the message just
  received along edge $e$}. This is important so that only
``extrinsic'' information is passed along from a node to its neighbor
in each step. It is exactly this restriction that leads to the
independence condition that makes analysis of the decoding possible.

In light of the above restriction, the iterative decoding can be
described in terms of the following message maps: $\Psi_v^{(\ell)} :
{\cal Y} \times {\cal M}^{d_v-1} \rightarrow {\cal M}$ for variable
node $v$ with degree $d_v$ for the $\ell$'th iteration, $\ell \ge 1$, and
$\Psi_c^{(\ell)} : {\cal M}^{d_v-1} \rightarrow {\cal M}$ for check
node $c$ with degree $d_c$.  Note the message maps can be different
for different iterations, though several powerful choices exist
where they remain the same for all iterations (and we will mostly
discuss such decoders). Also, while the message maps can be different
for different variable (and check) nodes, we will use the same map
(except for the obvious dependence on the degree, in case of irregular
graphs).

The intuitive interpretation of messages is the following. A message
is supposed to be an estimate or guess of a particular codeword
bit. For messages that take $\pm 1$ values, the guess on the bit is
simply the message itself. We can also add a third value, say $0$,
that would signify an erasure or abstention from guessing the value of
the bit. More generally, messages can take values in a larger discrete
domain, or even take continuous values. In these cases the sign of the
message is the estimated value of the codeword bit, and its absolute
value is a measure of the reliability or confidence in the estimated
bit value.

\subsection{Symmetry Assumptions}
We have already discussed the output-symmetry condition of the
channels we will be interested in, i.e., $p(y=q|x=1) = p(y=-q|x=-1)$. 
We now mention two reasonable symmetry assumptions on the message
maps, which will be satisfied by the message maps underlying the
decoders we discuss:
\begin{itemize}
\item {\bf Check node symmetry:} Signs factor out of check node
  message maps, i.e., for all $(b_1,\dots,b_{d_c-1}) \in \{1,-1\}^{d_c-1}$
\[ \Psi_c^{(\ell)}(b_1m_1,\cdots , b_{d_c-1} m_{d_c-1}) = \left(
  \prod_{i=1}^{d_c-1} b_i \right)
  \Psi_c^{(\ell)}(m_1,\cdots ,m_{d_c-1})  \ .\]
\item {\bf Variable node symmetry:} If the signs of all messages
  into a variable node are flipped, then the sign of its output gets
  flipped:
\[ \Psi_v^{(\ell)}(-m_0,-m_1,\cdots , - m_{d_v-1}) =
  -\Psi_v^{(\ell)}(m_0,m_1,\cdots ,m_{d_c-1})  \ .\]
\end{itemize}

When the above symmetry assumptions are fulfilled and the channel is
output-symmetric, the decoding error probability is independent of the
actual codeword transmitted.  Indeed, it is not hard (see, for
instance \cite[Lemma 1]{RU01}) to show that when a codeword
$(x_1,\dots,x_n)$ is transmitted and $(y_1,\dots,y_n)$ is received
where $y_i = x_i z_i$, the messages to and from the variable node
$v_i$ are equal to $x_i$ times the corresponding message when the
all-ones codeword is transmitted and $(z_1,\dots,z_n)$ is received.
Therefore, the entire behavior of the decoder can be predicted from
its behavior assuming transmission of the all-ones codeword (recall
that we are using $\{1,-1\}$ notation for the binary alphabet). So,
for the analysis, we will assume that the all-ones codeword was
transmitted.

\section{Regular LDPC codes and simple iterative decoders}
\label{sec:regular}
We will begin with regular LDPC codes and a theoretical analysis of simple
message-passing algorithms for decoding them.

\subsection{Gallager's program}
\label{sec:gallager-program}
The story of LDPC codes and iterative decoding begins in Gallager's
remarkable Ph.D. thesis completed in 1960, and later published in
1963~\cite{gallager}. Gallager analyzed the behavior of a code picked
randomly from the ensemble of $(d_v,d_c)$-regular LDPC codes of a
large block length. He proved that with high probability, as $d_v$ and
$d_c$ increase, the rate vs. minimum distance trade-off of the code
approaches the Gilbert-Varshamov bound. Gallager also analyzed the
error probability of maximum likelihood (ML) decoding of random
$(d_c,d_c)$-regular LDPC codes, and showed that LDPC codes are at
least as good on the BSC as the optimum code a somewhat higher rate
(refer to \cite{gallager} for formal details concerning this
statement). This demonstrated the promise of LDPC codes independently
of their decoding algorithms (since ML decoding is the optimal
decoding algorithm in terms of minimizing error probability).

To complement this statement, Gallager also proved a ``negative''
result showing that for each finite $d_c$, there is a finite gap to
capacity on the BSC when using regular LDPC codes with check node degrees $d_c$
More precisely, he proved that the largest rate that can be achieved
for $\bsc_p$ with error
probability going to zero is at most $1 - \frac{H(p)}{H(p_{d_c})}$ where
$p_{d_c} = \frac{1 + (1-2p)^{d_c}}{2}$.  This claim holds even for
irregular LDPC codes with $d_c$ interpreted as the maximum check node
degree. This shows that the maximum check node degree needs to grow with the
gap $\eps$ between the rate of the code and capacity of the BSC.

Since only exponential time solutions to the ML decoding problem are
known, Gallager also developed simple, iterative decoding algorithms
for LDPC codes. These form the precursor to the modern day
message-passing algorithms. More generally, he laid down the
foundations of the following program for determining the threshold
channel parameter below which a suitable LDPC code can be used in
conjunction with a given iterative decoder for reliable information
transmission.
\begin{description}
\item[Code construction:] Construct a family of $(d_v,d_c)$-regular
  factor graphs with $n$ variable nodes (for increasing $n$) with
  girth greater than $4 \ell(n) = \Omega(\log n)$. An explicit
  construction of such graphs was also given by
  Gallager~\cite[Appendix C]{gallager}.
\item[Analysis of Decoder:] Determine the average fraction of
  incorrect\footnote{A message is incorrect if the bit value it
  estimates is wrong. For transmission of the all-ones codeword, this
  means the message has a non-positive value.} messages passed at the
  $i$'th iteration of decoding for $i \le \ell = \ell(n)$ (assuming there are no cycles of
  length at most $4\ell$). This fraction is usually expressed by a
  system of recursive equations that depend on $d_v,d_c$ and the
  channel parameter (such as crossover probability, in case of the
  BSC).
\item[Threshold computation:] Using the above equations, compute
  (analytically or numerically) the threshold channel parameter below
  which the expected fraction of incorrect messages approaches zero as
  the number of iterations increases.  Conclude that the chosen
  decoder when applied to this family of codes with $\ell(n)$ decoding
  rounds leads to bit-error probability approaching zero as long as
  the channel parameter is below the threshold.
\end{description}

The recent research on (irregular) LDPC codes shares the same
essential features of the above program. The key difference is that
the requirement of an explicit code description in Step 1 is
relaxed. This is because for irregular graphs with specific
requirements on degree distribution, explicit constructions of large
girth graphs seem very hard. Instead, a factor graph chosen randomly
from a suitable ensemble is used. This raises issues such as the
concentration of the performance of a random code around the average
behavior of the ensemble. It also calls for justification of the large
girth assumption in the decoding. We will return to these aspects when
we begin our discussion of irregular LDPC codes in
Section~\ref{sec:irregular}. 

We should point out that Gallager himself used random regular LDPC
codes for his experiments with iterative decoders for various channels
such as the BSC, the BIAWGN, and the Rayleigh fading channel. However,
if we so desire, for the analytic results, even explicit constructions
are possible.  In the rest of this section, we assume an explicit
large girth factor graph is used, and focus on the analysis of some
simple and natural iterative decoders. Thus the only randomness
involved is the one realizing the channel noise.

\subsection{Decoding on the binary erasure channel}
\label{sec:bec-algo}
Although Gallager did not explicitly study the BEC, his methods
certainly apply to it, and we begin by studying the BEC.  For the BEC,
there is essentially a unique choice for a non-trivial message-passing
decoding algorithm. In a variable-to-check message round, a variable
whose bit value is known (either from the channel output or from a
check node in a previous round) passes along its value to the
neighboring check nodes, and a variable whose bit value is not yet
determined passes a symbol (say $0$) signifying erasure. In the
check-to-variable message round, a check node $c$ passes to a neighbor
$v$ an erasure if it receives an erasure from at least one neighbor
besides $v$, and otherwise passes the bit value $b$ to $v$ where $b$
is the parity of the bits received from neighbors other than $v$.
Formally, the message maps are given as follows:
\[ \Psi^{(\ell)}_v(r,m_1,\dots,m_{d_v-1}) = \left\{  \begin{array}{ll} 

b & \mbox{if at least
one of $r,m_1,\dots,m_{d_v-1}$ equals $b \in \{1,-1\}$} \\
0 & \mbox{if $r = m_1 = \cdots = m_{d_v-1}=0$} \\
 \end{array}
\right.
\]
(Note that the map is well-defined since the inputs to a variable node
will never give conflicting $\pm 1$ votes on its value.)
\[ \Psi^{(\ell)}_c(m_1,\dots,m_{d_c-1}) = \prod_{i=1}^{d_c-1} m_i \]

We note that an implementation of the decoder is possible that uses
each edge of the factor for message passing exactly once.  Indeed,
once a variable node's value is known, the bit value is communicated
to its neighboring check nodes, and this node (and edges incident on
it) are removed from the graph. Each check node maintains the parity
of the values received from its neighboring variables so far, and
updates this after each round of variable messages (note that it
receives each variable node's value exactly once). When a check node
has degree exactly one (i.e., values of all but one of its variable
node neighbors are now known), it communicates the parity value it has
stored to its remaining neighbor, and both the check node and the
remaining edge incident on it are deleted. This version of the
iterative decoder has been dubbed the {\em Peeling Decoder}.  The
running time of the Peeling Decoder is essentially the number of edges
in the factor graph, and hence it performs about $d_v$ operations per
codeword bit.

Let us analyze this decoding algorithm for $\ell$ iterations, where
$\ell$ is a constant (chosen large enough to achieve the desired
bit-error probability). We will assume that the factor graph does not
have any cycle of length at most $4\ell$ (which is certainly true if
it has $\Omega(\log n)$ girth). 

The following is crucial to our analysis.
\begin{lemma}
For each node, the random variables
corresponding to the messages received by it in the $i$'th iteration
are all independent, for $i \le \ell$.
\end{lemma}
Let us justify why the above is the
case. For this, we crucially use the fact that the message sent along
an edge, say from $v$ to $c$, does not depend on the
message that $v$ receives from $c$. Therefore, the information received
at a check node $c$ (the situation for variable nodes is identical)
from its neighbors in the $i$'th iteration is determined by
by a
computation graph rooted at $c$, with its $d_c$ variable node
neighbors as its children, the
$d_v-1$ neighbors besides $c$ of each these variable nodes as their
children, the $d_c-1$ other neighbors of these check nodes as their
children, and so on.  Since the girth of the graph is greater than
$4\ell$, the computation graph is in fact a tree. Therefore, the
information received by $c$ from its neighbors in the $i$'th iteration
are all independent.

Take an arbitrary edge $(v,c)$ between variable node $v$ and check node
$c$. Let us compute the probability $p_i$ that the message from $v$ to
$c$ in the $i$'th iteration is an erasure (using induction and the
argument below, one can justify the claim that this probability, which
is taken over the channel noise, will be independent of the edge and
only depend on the iteration number, as long as $i \le \ell$). For
$i=0$, $p_0 = \a$, the probability that the bit value for $v$ was
erased by the $\bec_\a$.  In the $(i+1)$'st iteration, $v$ passes an
erasure to $c$ iff it was originally erased by the channel, and it
received an erasure from each of its $d_v-1$ neighbors other than
$c$. Each of these neighboring check nodes $c'$ in turn sends an
erasure to $v$ iff at least one neighbor of $c'$ other than $v$ sent
an erasure to $c'$ during iteration $i$ --- due to the independence of
the involved messages, this event occurs for node $c'$ with probability
$(1-(1-p_i)^{d_c-1})$. Again, because the messages from various check
nodes to $v$ in the $(i+1)$'st round are independent, we have
\begin{equation}
\label{eq:bec-recursion}
p_{i+1} = \a \cdot (1-(1-p_i)^{d_c-1})^{d_v-1}  \ . 
\end{equation}

By linearity of expectation, $p_i$ is the expected fraction of
variable-to-check messages sent in the $i$'th iteration that are
erasures. We would like to show that $\lim_{\ell \to \infty} p_{\ell}
= 0$, so that the bit-error probability of the decoding vanishes as
the number of iterations grows. The largest erasure probability $\a$
for which this happens is given by the following lemma.
\begin{lemma}
\label{lem:bec-reg-threshold}
The threshold erasure probability $\a^{\bp}(d_v,d_c)$ for the BEC below which the
message-passing algorithm results in vanishing bit-erasure probability
is given by
\begin{equation}
\label{eq:bec-threshold}
\a^{\bp}(d_v,d_c) = \min_{x \in [0,1]}
\frac{x}{(1-(1-x)^{d_c-1})^{d_v-1}}  \ . 
\end{equation}
\end{lemma}
\begin{proof} By definition, $\a^{\bp}(d_v,d_c)
  = \sup \{ \a \in [0,1] : \lim_{i \to \infty} p_i = 0\}$ where $p_i$
  is as defined recursively in (\ref{eq:bec-recursion}). Define the
  functions $g(x)= \frac{x}{(1-(1-x)^{d_c-1})^{d_v-1}}$, and $f(\a,x)
  = \a (1-(1-x)^{d_c-1})^{d_v-1}$.  Also let $\a^* = \min_{x \in
  [0,1]} g(x)$. We wish to prove that $\a^{\bp}(d_v,d_c) =\a^*$.

If $\a < \a^*$, then for every $x \in [0,1]$, $f(\a,x) = \frac{\a
  x}{g(x)} \le \frac{\a^* x}{g(x)} \le x$, and in fact $f(\a,x) < x$
  for $x \in (0,1]$. Hence it follows that $p_{i+1} = f(\a,p_i) \le
  p_i$ and since $0 \le f(\a,x) \le \a$ for all $x \in [0,1]$, the
  probability converges to a value $p_\infty \in [0,\a]$. Since $f$ is
  continuous, we have $p_\infty = f(\a,p_\infty)$, which implies
  $p_\infty = 0$ (since $f(\a,x) < x$ for $x > 0$). This shows that
  $\a^{\bp}(d_v,d_c) \ge \a^*$.

Conversely, if $\a > \a^*$, then let $x_0 \in [0,1]$ be such that $\a
> g(x_0)$. Then $\a \ge f(\a,x_0) = \frac{\a x_0}{g(x_0)} > x_0$, and
of course $f(\a,\a) \le \a$. Since $f(\a,x)$ is a continuous function of
$x$, we must have $f(\a,x^*) = x^*$ for some $x^* \in (x_0,\a]$. For
the recursion (\ref{eq:bec-recursion}) with a fixed value of $\a$, it is easy to see by induction
that if
$p_0 \ge  p'_0$, then $p_i \ge p'_i$ for all $i \ge 1$. If $p'_0 = x^*$,
then we have $p'_i = x^*$ for all $i$. Therefore, when $p_0 = \a \ge
x^*$, we have $p_i \ge x^*$ for all $i$ as well. In other words, the
error probability stays bounded below by $x^*$ irrespective of the
number of iterations. This proves that $\a^{\bp}(d_v,d_c) \le \a^*$.

Together, we have exactly determined the threshold to be $\a^* = \min_{x \in
  [0,1]} g(x)$. 
\end{proof}

\begin{remark}
Using standard calculus, we can determine $\a^{\bp}(d_v,d_c)$ to be
$\frac{1-\gamma}{(1-\gamma^{d_c-1})^{d_v-1}}$ where $\gamma$ is the
unique positive root of the polynomial $p(x) =
((d_v-1)(d_c-1)-1)x^{d_c-2} - \sum_{i=0}^{d_c-3} x^i$.  Note that when
$d_v = 2$, $p(1) = 0$, so the threshold equals $0$. Thus we must pick
$d_v \ge 3$, and hence $d_c \ge 4$ (to have positive rate). For the
choice $d_v=3$ and $d_c=4$, $p(x)$ is a quadratic and we can
analytically compute $\a^{\bp}(3,4) \approx 0.6474$; note that
capacity for this rate equals $3/4=0.75$.  (The best threshold one can
hope for equals $d_v/d_c$ since the rate is at least
$1-d_v/d_c$.)  Closed form analytic expressions for some other small
values of $(d_v,d_c)$ are given in \cite{BRU04}: for example,
$\a^{\bp}(3,5) \approx 0.5406$ (compare to capacity of $0.6$) and
$\a^{\bp}(3,6) \approx 0.4294$ (compare to capacity of $0.5$).
\end{remark}

\begin{theorem}
For integers $3 \le d_v < d_c$, there exists an explicit family of
binary linear codes of rate at least $1-\frac{d_v}{d_c}$ that can be
reliably decoded in linear time on $\bec_{\a}$ provided $\a <
\a^{\bp}(d_v,d_c)$.\footnote{Our analysis showed that the bit-error
  probability can be made below any desired $\eps > 0$ by picking the
  number of iterations to be a large enough constant. A more careful
  analysis using $\ell(n) = \Omega(\log n)$ iterations shows that
  bit-error probability is at most $\exp(-n^{\beta})$ for some
  constant $\beta = \beta(d_v,d_c)$. By a union bound, the entire
  codeword is thus correctly recovered with high probability.}
\end{theorem}

\subsection{Decoding on the BSC}
%: Gallagher's ``Algorithm A'' and extensions}

The relatively clean analysis of regular LDPC codes on the BEC is
surely encouraging. As mentioned earlier, Gallager in fact did not
consider the BEC in his work. We now discuss one of his decoding
algorithms for the BSC, that has been dubbed Gallager's Algorithm A,
and some simple extensions of it.
 
\subsubsection{Gallager's Algorithm A}
The message alphabet of Algorithm A will equal $\{1,-1\}$, so the
nodes simply pass guesses on codeword bits. The message maps are time
invariant and do not depend on the iteration number, so we will omit
the superscript indicating the iteration number in describing the
message maps. The check nodes
send a message to a variable node indicating the parity of the {\em other}
neighboring variables, or formally:
\[ \Psi_c(m_1,\dots,m_{d_c-1}) = \prod_{i=1}^{d_c-1} m_i \ . \]
The variable nodes send to a neighboring check node their original received value unless the
incoming messages from the {\em other} check nodes unanimously
indicate otherwise, in which case it sends the negative of the
received value. Formally,
\[
\Psi_v(r,m_1,\dots,m_{d_v-1}) = \left\{  \begin{array}{ll} 
-r & \mbox{if  $m_1 = \cdots = m_{d_v-1} = -r$} \\
r & \mbox{otherwise} \ .
 \end{array}
\right.
\]
As in the case of BEC, we will track the expected fraction of
variable-to-check node messages that are erroneous in the $i$'th
iteration.  Since we assume the all-ones codeword was transmitted,
this is simply the expected fraction of messages that equal $-1$. Let
$p_i$ be the probability (over the channel noise) that a particular
variable-to-check node message in iteration $i$ equals $-1$ (as in the
case of the BEC, this is independent of the actual edge for $i \le
\ell$). Note that we have $p_0 = p$, the crossover probability of the
BSC.

It is a routine calculation using the independence of the incoming
messages to prove the following recursive
equation~\cite[Sec. 4.3]{gallager}, \cite[Sec III]{RU01}:
\begin{equation}
p_{i+1} = p_0 - p_0 \left(\frac{1+(1-2p_i)^{d_c-1}}{2}\right)^{d_v-1} + (1-p_0) \left(\frac{1-(1-2p_i)^{d_c-1}}{2}\right)^{d_v-1}
\end{equation}
For a fixed value of $p_0$, $p_{i+1}$ is a increasing function of
$p_i$, and for a fixed value of $p_i$, $p_{i+1}$ is an increasing
function of $p_0$. Therefore, by induction $p_i$ is an increasing
function of $p_0$. Define the threshold value of this algorithm ``A''
as $p^{A}(d_v,d_c) = \sup \{ p_0 \in [0,1] : \lim_{\ell \to \infty}
p_\ell = 0\}$. By the above argument, if the crossover probability $p
< p^{A}(d_v,d_c)$, then the expected fraction of erroneous messages in
the $\ell$'th iteration approaches $0$ as $\ell \to \infty$. 

Regardless of the exact quantitative value, we want to point out that
when $d_v \ge 3$, the threshold is positive. Indeed, for $d_v > 2$,
for small enough $p_0 > 0$, one can see that $p_{i+1} < p_i$ for $0 <
p_i \le p_0$ and $p_{i+1} = p_i$ for $p_i =0$, which means that
$\lim_{i \to \infty} p_i = 0$.

Exact analytic expressions for the threshold have been computed for
some special cases~\cite{BRU04}. This is based on the characterization
of $p^{A}(d_v,d_c)$ as the supremum of all $p_0 > 0$ for which
\[ x = p_0 - p_0 \left(\frac{1+(1-2x)^{d_c-1}}{2}\right)^{d_v-1} + (1-p_0) \left(\frac{1-(1-2x)^{d_c-1}}{2}\right)^{d_v-1} \]
does not have a strictly positive solution $x$ with $x \le p_0$.
Below are some example values of the threshold (up to the stated
precision). Note that the rate of the code is $1-d_v/d_c$ and the
Shannon limit is $H^{-1}(d_v/d_c)$ (where $H^{-1}(y)$ for $0 \le y \le
1$ is defined as the unique value of $x \in [0,1/2]$ such that
$H(x)=y$).

\begin{tabular}{|l|l|l|l|}
\hline 
$d_v$ & $d_c$ & $p^{A}(d_v,d_c)$ & Capacity \\
\hline 
3 & 6 & 0.0395 & 0.11 \\
4 & 8 & 1/21 & 0.11 \\
5 & 10 & 1/36 & 0.11 \\
4 & 6 & 1/15 & 0.174 \\
3 & 4 & 0.106 & 0.215 \\
3 & 5 & 0.0612 & 0.146 \\
\hline 
\end{tabular}

\subsubsection{Gallager's Algorithm B}
\label{sec:gallager-algoB}
Gallager proposed an extension to the above algorithm, which is now
called Gallager's Algorithm B, in which a variable node decides to
flip its value in an outgoing message when at least $b$ of the
incoming messages suggest that it ought to flip its value. In
Algorithm A, we have $b = d_v-1$. The threshold $b$ can also depend on
the iteration number, and we will denote by $b_i$ this value during
the $i$'th iteration. Formally, the variable message map in the $i$'th
iteration is given by
\[
\Psi^{(i)}_v(r,m_1,\dots,m_{d_v-1}) = \left\{  \begin{array}{ll} 
-r & \mbox{if  $| \{ j : m_j = -r\}| \ge b_i$} \\
r & \mbox{otherwise} \ .
 \end{array}
\right.
\]
The check node message maps remain the same. 
The threshold should be greater than $(d_v-1)/2$ since intuitively one
should flip only when more check nodes suggest a flip than those that
suggest the received value. So when $d_v = 3$, the above algorithm
reduces to Algorithm A.

Defining the probability of an incorrect variable-to-check node
message in the $i$'th iteration to be $\tilde{p}_i$, one can show the
recurrence~\cite[Sec. 4.3]{gallager}:
\begin{align*}
\tilde{p}_{i+1} & =  \tilde{p}_0 - \tilde{p}_0 \sum_{j=b_{i+1}}^{d_v-1}
      {\textstyle {{d_v-1} \choose j}}
      \left(\frac{1+(1-2\tilde{p}_i)^{d_c-1}}{2}\right)^j
\left(\frac{1-(1-2\tilde{p}_i)^{d_c-1}}{2}\right)^{d_v-1-j}  
\\ 
& +
	  (1-\tilde{p}_0)   \sum_{j=b_{i+1}}^{d_v-1}
 {\textstyle {{d_v-1} \choose j}}
  \left(\frac{1+(1-2\tilde{p}_i)^{d_c-1}}{2}\right)^{d_v-1-j}
\left(\frac{1-(1-2\tilde{p}_i)^{d_c-1}}{2}\right)^{j} 
\end{align*}
The cut-off value $b_{i+1}$ can then be chosen to minimize this
value. The solution to this minimization is the smallest integer $b_{i+1}$
for which
\[ \frac{1-\tilde{p}_0}{\tilde{p}_0} \le
\left(
\frac{1+(1-2\tilde{p}_i)^{d_c-1}}{1-(1-2\tilde{p}_i)^{d_c-1}}\right)^{2b_{i+1}-d_v+1}
\ . \] By the above expression, we see that as $\tilde{p}_i$
decreases, $b_{i+1}$ never increases. And, as $\tilde{p}_i$ is sufficiently
small, $b_{i+1}$ takes the value $d_v/2$ for even $d_v$ and
$(d_v+1)/2$ for odd $d_v$. Therefore, a variable node flips its value
when a majority of the $d_v-1$ incoming messages suggest that the
received value was an error. We note that this majority criterion for
  flipping a variable node's bit value was
 also used in decoding of expander codes~\cite{SS96}.

Similar to the analysis of Algorithm A, using the above recurrence, one can show that when
$d_v \ge 3$, for sufficiently small $p_0 > 0$, we have $p_{i+1} < p_i$
when $0 < p_i \le p_0$, and of course when $p_i=0$, we have $p_{i+1} =
0$. Therefore, when $d_v \ge 3$, for small enough $p_0 > 0$, we have
$\lim_{i \to \infty} p_i = 0$ and thus a positive threshold.

The values of the threshold of this algorithm for small pairs
$(d_v,d_c)$ appear in \cite{RU01}. For the pairs $(4,8)$, $(4,6)$ and
$(5,10)$ the thresholds are about $0.051$, $0.074$, and $0.041$
respectively.  For comparison, for these pairs Algorithm A achieved a
threshold of about $0.047$, $0.066$, and $0.027$ respectively.

\subsubsection{Using Erasures in the Decoder}
In both the above algorithms, each message made up its mind on whether
to guess $1$ or $-1$ for a bit. But it may be judicious to sometimes
abstain from guessing, i.e., to send an ``erasure'' message (with value
$0$), if there is no good reason to guess one way or the other. For
example, this may be the appropriate course of action if a variable
node receives one-half $1$'s and one-half $-1$'s in the incoming check
node messages. This motivates an algorithm with message alphabet
$\{1,0,-1\}$ and the following message maps (in iteration $\ell$):
\[ \Psi^{(\ell)}_v(r,m_1,m_2,\dots,m_{d_v-1}) = \mathsf{sgn} \left( w^{(\ell)} r +
\sum_{j=1}^{d_v-1} m_j \right) \]
and 
\[ \Psi^{(\ell)}_c(m_1,m_2,\dots,m_{d_c-1}) = \prod_{j=1}^{d_c-1} m_j
\ . \]
The weight $w^{(\ell)}$ dictates the relative importance given to the
received value compared to the suggestions by the check nodes in the
$\ell$'th iteration.  These weights add another dimension of design choices
that one can optimize.

Exact expressions for the probabilities $p^{(-1)}_i$ and $p^{(0)}_i$)
that a variable-to-check message is an error (equals $-1$) and an
erasure (equals $0$) respectively in the $i$'th iteration can be
written down~\cite{RU01}. These can be used to pick appropriate
weights $w^{(i)}$. For the $(3,6)$-regular code,
$w^{(1)}=2$ and $w^{(i)} = 1$ for $i \ge 2$ is reported as the optimum
choice in \cite{RU01}, and using this choice the resulting algorithm
has a threshold of about $0.07$, which is a good improvement over the
$0.04$ achieved by Algorithm A. More impressively, this is close to
the threshold of $0.084$ achieves by the ``optimal'' belief
propagation decoder. A heuristic to pick the weights $w^{(i)}$ is
suggested in \cite{RU01} and the threshold of the resulting algorithm
is computed for small values of $(d_v,d_c)$. 

\subsection{Decoding on BIAWGN}

We now briefly turn to the BIAWGN channel. We discussed the most
obvious quantization of the channel output which converts the channel
to a BSC with crossover probability $Q(1/\sigma)$. There is a natural
way to incorporate erasures into the quantization. We pick a threshold
$\tau$ around zero, and quantize the AWGN channel output $r$ into
$-1$, $0$ (which corresponds to erasure), or $1$ depending on whether
$r \le -\tau$, $-\tau < r < \tau$, or $r \ge \tau$, respectively. We
can then run exactly the above message-passing algorithm (the one
using erasures). More generally, we can pick a separate threshold
$\tau_i$ for each iteration $i$ --- the choice of $\tau_i$ and
$w^{(i)}$ can be optimized using some heuristic criteria. Using this
approach, a threshold of $\sigma^* = 0.743$ is reported for
communication using a $(3,6)$-regular LDPC code on the BIAWGN
channel. This corresponds to a raw bit-error probability of
$Q(1/\sigma^*) = 0.089$, which is almost $2\%$ greater than the
threshold crossover probability of about $0.07$ achieved on the
BSC. So even with a ternary message alphabet, providing soft
information (instead of quantized hard bit decisions) at the input to
the decoder can be lead to a good performance gain. The belief
propagation algorithm we discuss next uses a much large message
alphabet and yields further substantial improvements for the BIAWGN.

\subsection{The belief propagation decoder}

So far we have discussed decoders with quantized, discrete messages
taking on very few values. Naturally, we can expect more powerful
decoders if more detailed information, such as real values quantifying
the likelihood of a bit being $\pm 1$, are passed in each iteration.
We now describe the ``belief propagation'' (BP) decoder which is an
instance of such a decoder (using a continuous message alphabet). We
follow the description in \cite[Sec. III-B]{RU01}. In
belief propagation, the messages sent along an edge $e$ represent the
posterior conditional distribution on the bit associated with
the variable node incident on $e$. This distribution corresponds to a pair of
nonnegative reals $p_1,p_{-1}$ satisfying $p_1+p_{-1} = 1$. This
pair can be encoded as a single real number (including $\pm \infty$)
using the log-likelihood ratio $\log \frac{p_1}{p_{-1}}$, and the
messages used by the BP decoder will follow this representation.

Each node acts under the assumption that each message communicated to
it in a given round is a conditional distribution on the associated
bit, and further each such message is conditionally independent of the
others. Upon receiving the messages, a node transmits to each neighbor
the conditional distribution of the bit conditioned on all information
{\em except} the information from that neighbor (i.e., only extrinsic
information is used in computing a message). If the graph has large
enough girth compared to the number of iterations, this assumption is
indeed met, and the messages at each iteration reflect the true
log-likelihood ratio given the observed values in the tree
neighborhood of appropriate depth.

If $l_1,l_2,\dots,l_k$ are the likelihood ratios of the conditional
distribution of a bit conditioned on independent random variables,
then the likelihood ratio of the bit value conditioned on all of the
random variables equals $\prod_{i=1}^k l_i$. Therefore,
log-likelihoods of independent messages add up, and this leads to the
variable message map (which is independent of the iteration number):
\[ \Psi_v(m_0,m_1,\dots,m_{d_v-1}) = \sum_{i=0}^{d_v-1} 
m_i \]
where $m_0$ is the log-likelihood ratio of the bit based on the
received value (eg., for the $\bsc_p$, $m_0 = r \log \frac{1-p}{p}$ where
$r \in \{1,-1\}$ is the received value).

The performance of the decoder is analyzed by tracking the evolution
of the probability density of the log-likelihood ratios (hence the
name ``density evolution'' for this style of analysis). By the above,
given densities $P_0,P_1,\dots,P_{d_v-1}$ on the real quantities
$m_0,m_1,\dots,m_{d_v-1}$, the density of
$\Psi_v(m_0,m_1,\dots,m_{d_v-1})$ is the convolution 
$P_0 \otimes P_1 \otimes \cdots \otimes P_{d_v-1}$
over the reals of those densities.  In the computation, one has
$P_1=P_2=\cdots=P_{d_v-1}$ and the densities will be quantized, and
the convolution can be efficiently computed using the FFT.

Let us now turn to the situation for check nodes. Given bits $b_i$, $1
\le i \le k$, with independent
probability distributions $(p_1^i,p_{-1}^i)$, what is the distribution
$(p_1,p_{-1})$ of the bit $b = \prod_{i=1}^k b_i$? We have the
expectation 
\[ E[b] = E [\prod_i b_i] = \prod_i E[b_i] = \prod_i (p_1^i
- p_{-1}^i) \ . \] Therefore we have $p_1-p_{-1} = \prod_{i=1}^k
  (p_1^i - p_{-1}^i)$. Now if $m$ is the log-likelihood ratio
  $\log\frac{p_1}{p_{-1}}$, then $p_1-p_{-1} = \frac{e^m-1}{e^m+1} =
  \tanh (m/2)$. Conversely, if $p_1-p_{-1} = q$,
  then $\log \frac{p_1}{p_{-1}} = \log \frac{1+q}{1-q}$. These
  calculations lead to the following check node map for the
  log-likelihood ratio:
\[ \Psi_c(m_1,m_2,\dots,m_{d_c-1}) = \log \left( \frac{
 1+
  \prod_{i=1}^{d_c-1} \tanh (m_i/2)}
{ 1 -
  \prod_{i=1}^{d_c-1} \tanh (m_i/2)} \right) \ . \]
It seems complicated to track the density of
$\Psi_c(m_1,m_2,\dots,m_{d_c-1})$ based on those of the
$m_i$'s. However, as shown in \cite{RU01}, this can be also be
realized via a Fourier transform, albeit with a slight change in
representation of the conditional probabilities $(p_1,p_{-1})$. We
skip the details and instead point the reader to
\cite[Sec. III-B]{RU01}.

Using these ideas, we have an effective algorithm to recursively
compute, to any desired degree of accuracy, the probability density
$P^{(\ell)}$ of the log-likelihood ratio of the variable-to-check node
messages in the $\ell$-th iteration, starting with an explicit
description of the initial density $P^{(0)}$. The initial density is
simply the density of the log-likelihood ratio of the received value,
assuming transmission of the all-ones codeword; for example, for
$\bsc_p$, the initial density $P^{(0)}$ is given by
\[ P^{(0)}(x) = p \delta\left(x - \log \frac{p}{1-p} \right) 
+ (1-p) \delta \left( x - \log \frac{1-p}{p} \right) \ , \]
where $\delta(x)$ is the Dirac delta function.

The threshold crossover probability for the BSC and the threshold
variance for the BIAWGN under belief propagation decoding for various
small values of $(d_v,d_c)$ are computed by this method and reported
in \cite{RU01}. For the $(3,6)$ LDPC code, these thresholds are
respectively $p^* =0.084$ (compare with Shannon limit of $0.11$) and
$\sigma^* = 0.88$ (compare with Shannon limit of $0.9787$).

The above numerical procedure for tracking the evolution of densities for
belief propagation and computing the associated threshold to any
desired degree of accuracy has since been applied with great
success. In \cite{RSU01}, the authors apply this method to irregular
LDPC codes with optimized structure and achieve a threshold of
$\sigma^* = 0.9718$ with rate $1/2$ for the BIAWGN, which is a mere
$0.06$ dB way from the Shannon capacity limit.\footnote{The threshold
  signal-to-noise ratio $1/(\sigma^*)^2 = 0.2487$ dB, and the Shannon
  limit for rate $1/2$ is $0.187$ dB.}

\section{Irregular LDPC codes}
\label{sec:irregular}

Interest in LDPC codes surged following the seminal paper~\cite{LMSS}
that initiated the study of irregular LDPC codes, and proved their
potential by achieving the capacity on the BEC. Soon, it was realized
that the benefits of irregular LDPC codes extend to more powerful
channels, and this led to a flurry of activity. In this section, we
describe some of the key elements of the analytic approach used to
to study message-passing decoding algorithms for irregular LDPC
codes.

\subsection{Intuitive benefits of irregularity}

We begin with some intuition on why one might expect improved
performance by using irregular graphs. In terms of iterative decoding,
from the variable node perspective, it seems better to have high
degree, since the more information it gets from check nodes, the more
accurately it can guess its correct value. On the other hand, from the
check node perspective, the lower its degree, the more valuable the
information it can transmit back to its neighbors. (The XOR of several
mildly unpredictable bits has a much larger unpredictability.) But in
order to have good rate, there should be far fewer check nodes than
variable nodes, and therefore meeting the above competing requirements
is challenging. Irregular graphs provide significantly more flexibility
in balancing the above incompatible degree requirements. It seems
reasonable to believe that a wide spread of degrees for
variable nodes could be useful. This is because one might expect
that variable nodes with high degree will converge to their correct
value quickly. They can then provide good information to the
neighboring check nodes, which in turn provide better information to
lower degree variable nodes, and so on leading to a cascaded wave effect.

The big challenge is to leap from this intuition to the design of
appropriate irregular graphs where this phenomenon provably occurs,
and to provide analytic bounds on the performance of natural iterative
decoders on such irregular graphs.

Compared to the regular case, there are additional technical issues
revolving around how irregular graphs are parameterized, how they are
constructed (sampled), and how one deals with the lack of explicit
large-girth constructions. We discuss these issues in the next two
subsections.

\subsection{The underlying ensembles}

We now describe how irregular LDPC codes can be parameterized and
constructed (or rather sampled). Assume we have an LDPC code with $n$
variable nodes with $\Lambda_i$ variable nodes of degree $i$ and $P_i$
check nodes of degree $i$. We have $\sum_i \Lambda_i = n$, and $\sum_i
i \Lambda_i = \sum_i i P_i$ as both these equal the number of edges in
the graph. Also $\sum_i P_i = n(1-r)$ where $r$ is the designed rate
of the code.  It is convenient to capture this information in the
compact polynomial notation:
\[ \Lambda(x) = \sum_{i=2}^{d_v^{\max}} \Lambda_i x^i \ , \qquad P(x) = \sum_{i=1}^{d_c^{\max}} P_i x^i \ . \]
We call the polynomials $\Lambda$ and $P$ the variable and check
degree distributions from a node perspective. Note that $\Lambda(1)$ is the number of variable nodes, $P(1)$ the number of check nodes, and $\Lambda'(1) = P'(1)$ the number of edges. 

Given such a degree distribution pair $(\Lambda,P)$, let
$\ldpc(\Lambda,P)$ denote the ``standard'' ensemble of bipartite
(multi)graphs with $\Lambda(1)$ variable nodes and $P(1)$ check nodes,
with $\Lambda_i$ variable nodes and $P_i$ check nodes of degree $i$.
This ensemble is defined by taking $\Lambda'(1) = P'(1)$ ``sockets''
on each side, allocating $i$ sockets to a node of degree $i$ in some
arbitrary manner, and then picking a random matching between the
sockets.

To each member of $\ldpc(\Lambda,P)$, we associate the code of which
it is the factor graph. A slight technicality: since we are dealing
with multigraphs, in the parity check matrix, we place a non-zero
entry at row $i$ and column $j$ iff the $i$th check node is connected
to the $j$th variable node an {\em odd} number of times. Therefore, we
can think of the above as an ensemble of codes, and by abuse of
notation also refer to it as $\ldpc(\Lambda,P)$. (Note that the graphs
have a uniform probability distribution, but the induced codes need
not.) In the sequel, our LDPC codes will be obtained by drawing a
random element from the ensemble $\ldpc(\Lambda,P)$.

To construct a family of codes, one can imagine using a normalized
degree distribution giving the {\em fraction} of nodes of a certain
degree, and then considering an increasing number of nodes. For purposes of analysis, it ends up being convenient to use
normalized degree distributions from the {\em edge} perspective.  Let
$\lambda_i$ and $\rho_i$ denote the fraction of {\em edges} incident
to variable nodes and check nodes of degree $i$ respectively. That is,
$\lambda_i$ (resp. $\rho_i$) is the probability that a randomly chosen
edge is connected to a variable (resp. check) node of degree
$i$. These distributions can be compactly written in terms of the
power series defined below:
\[ \lambda(x) = \sum_{i} \lambda_i x^{i-1} \ , \qquad \rho(x) =\sum_i \rho_i x^{i-1} \ . \]
It is easily seen that $\lambda(x) = \frac{\Lambda'(x)}{\Lambda'(1)}$
and $\rho(x) = \frac{P'(x)}{P'(1)}$. 
If $M$ is the total number of edges, then the number of variable nodes
of degree $i$ equals $M\lambda_i/i$, and thus the total number of
variable nodes is $M \sum_i \lambda_i/i$. It follows that that the average
variable node degree equals $\frac{1}{\sum_i \lambda_i/i} =
\frac{1}{\int_0^1 \lambda(z)  dz}$. Likewise, the average check node
degree equals $\frac{1}{\int_0^1 \rho(z)  dz}$. It follows that the
designed rate can be expressed in terms of $\lambda,\rho$ as 
\begin{equation}
\label{eq:irreg-rate}
r = r(\lambda,\rho) = 1 - \frac{\int_0^1 \rho(z) dz}{\int_0^1
  \lambda(z) dz} \ . 
\end{equation}
We also have the inverse relationships
\begin{equation}
\label{eq:inverse-rel}
\frac{\Lambda(x)}{n} = \frac{\int_0^x \lambda(z)  dz}{\int_0^1
  \lambda(z) dz}\  ,  \quad \frac{P(x)}{n(1-r)} = \frac{\int_0^x
  \rho(z)  dz}{\int_0^1 \rho(z) dz}  \ .
\end{equation}
Therefore, $(\Lambda,P)$ and $(n,\lambda, \rho)$ carry the same
  information (in the sense we can obtain each from the other). For the asymptotic analysis we use
  $(n,\lambda,\rho)$ to refer to the LDPC code ensemble. There is a
  slight technicality that for some $n$, the $(\Lambda,P)$
  corresponding to $(n,\lambda,\rho)$ may not be integral. In this
  case, rounding the individual node distributions to the closest
  integer has negligible effect on the asymptotic performance of
  decoder or the rate, and so this annoyance may be safely ignored.

The degree distributions $\lambda,\rho$ play a prominent role in the
line of work, and the performance of the decoder is analyzed and
quantified in terms of these.

\subsection{Concentration around average performance}

Given a degree distribution pair $(\lambda,\rho)$ and a block length $n$,
the goal is to mimic Gallager's program (outlined in
Section~\ref{sec:gallager-program}), using a factor graph with degree
distribution $(\lambda,\rho)$ in place of a
$(d_v,d_c)$-regular factor graph. However, 
the task of constructing explicit
large girth graphs obeying precise irregular degree distributions
seems extremely difficult. Therefore, a key difference is to give up
on explicitness, and rather sample an element from the ensemble
$\ldpc(n,\lambda,\rho)$, which can be done easily as mentioned above.

It is not very difficult to show that a random code drawn from the
ensemble will have the needed girth (and thus be tree-like in a local
neighborhood of every edge/vertex) with high probability; see for
instance \cite[Appendix A]{RU01}. A more delicate issue is the
following: For the irregular case the neighborhood trees out of
different nodes have a variety of different possible structures, and
thus analyzing the behavior of the decoder on a specific factor graph
(after it has been sampled, even conditioning on it having large
girth) seems hopeless. What {\em is} feasible, however, is to analyze
the {\em average} behavior of the decoder (such as the expected fraction,
say $P_n^{(\lambda,\rho)}(\ell)$, of erroneous variable-to-check
messages in the $\ell$'th iteration) taken over all instances of the
code drawn from the ensemble $\ldpc(n,\lambda, \rho)$ and the
realization of the channel noise. It can be shown that, as $n \to
\infty$, $P_n^{(\lambda,\rho)}(\ell)$ converges to a certain quantity
$P_{\cal T}^{(\lambda,\rho)}(\ell)$, which is defined as the
probability (taken over both choice of the graph and the noise) that
an incorrect message is sent in the $\ell$'th iteration along an edge
$(v,c)$ assuming that the depth $2\ell$ neighborhood out of $v$ is a
tree. 

In order to define the probability $P_{\cal T}^{(\lambda,\rho)}(\ell)$
more precisely, one uses a ``tree ensemble'' ${\cal
T}_{\ell}(\lambda,\rho)$ defined inductively as follows. ${\cal
T}_0(\lambda,\rho)$ consists of the trivial tree consisting of just a
root variable node. For $\ell \ge 1$, to sample from ${\cal
T}_{\ell}(\lambda,\rho)$, first sample an element from ${\cal
T}_{\ell-1}(\lambda,\rho)$. Next for each variable leaf node
(independently), with probability $\lambda_{i+1}$ attach $i$ check
node children. Finally, for each of the new check leaf nodes,
independently attach $i$ variable node children with probability
$\rho_{i+1}$. The quantity $P_{\cal T}^{(\lambda,\rho)}(\ell)$ is then
formally defined as the probability that the outgoing message from the
root node of a sample $T$ from ${\cal T}_{\ell}(\lambda,\rho)$ is
incorrect, assuming the variable nodes are initially labeled with $1$
and then the channel noise acts on them independently (the probability
is thus both over the channel noise and the choice of the sample $T$
from ${\cal T}_{\ell}(\lambda,\rho)$).

The convergence of $P_n^{(\lambda,\rho)}(\ell)$ to $P_{\cal
T}^{(\lambda,\rho)}(\ell)$ is a simple consequence of the fact that,
for a random choice of the factor graph from $\ldpc(n,\lambda,\rho)$,
the depth $2\ell$ neighborhood of an edge is tree-like with
probability tending to $1$ as $n$ gets larger (for more details, see
\cite[Thm. 2]{RU01}).

The quantity $P_{\cal T}^{(\lambda,\rho)}(\ell)$ for the case of trees
is easily computed, similar to the case of regular graphs, by a
recursive procedure. One can then determine the threshold
channel parameter for which $P_{\cal T}^{(\lambda,\rho)}(\ell) \to 0$
as $\ell \to \infty$.

However, this only analyzed the {\em average}
behavior of the ensemble of codes. What we would like is for a random
code drawn from the ensemble $\ldpc(n,\lambda,\rho)$ to concentrate
around the average behavior with high probability. This would mean
that almost all codes behave alike and thus the individual behavior of
almost all codes is characterized by the average behavior of the
ensemble (which can be computed as outlined above).
A major success of this theory is that such a concentration phenomenon indeed holds, as shown in
\cite{LMSS-errors} and later extended to a large class of channels in
\cite{RU01}. The proof uses martingale arguments where the edges of
the factor graph and then the inputs to the decoder are revealed one
by one. We refrain from
presenting the details here and point the reader to
\cite[Thm. 1]{LMSS-errors} and \cite[Thm. 2]{RU01} (the result is
proved for regular ensembles in these works but extends to irregular
ensembles as long as the degrees in the graph are bounded).

In summary, it suffices to analyze and bound $P_{\cal
  T}^{(\lambda,\rho)}(\ell)$, and if this tends to $0$ as $\ell \to
  \infty$, then in the limit of a large number of decoding iterations,
  for almost all codes in the ensemble, the actual bit error
  probability of the decoder tends to zero for large enough block
  lengths.

\medskip \noindent {\bf Order of limits:} A remark on the order of the
  limits might be in order. The proposed style of analysis aims to
  determine the threshold channel parameter for which $\lim_{\ell \to
  \infty} \lim_{n \to \infty} E[P^{(\lambda,\rho)}_n(\ell)] = 0$. That
  is, we first fix the number of iterations and determine the limiting
  performance of an ensemble as the block length tends to infinity,
  and then let the number of iterations tend to infinity. Exchanging
  the order of limits gives us the quantity $\lim_{\ell \to \infty}
  \lim_{n \to \infty} E[P^{(\lambda,\rho)}_n(\ell)]$. It is this limit
  that corresponds to the more typical scenario in practice where for
  each fixed block length, we let the iterative decoder run until no
  further progress is achieved. We are then interested in the limiting
  performance as the block length tends to infinity.  For the BEC, it
  has been shown that for both the orders of taking limits, we get the
  same threshold~\cite[Sec. 2.9.8]{RU-book}. Based on empirical
  observations, the same has been conjectured for channels such as the
  BSC, but a proof of this seems to be out of sight.

\subsection{Analysis of average performance for the BEC}

We now turn to analyzing the average behavior of the ensemble
$\ldpc(n,\lambda,\rho)$ under message-passing decoding on the
BEC. (The algorithm for regular codes from Section~\ref{sec:bec-algo}
extends to irregular codes in the obvious fashion --- the
message maps are the same except the maps at different nodes will have different
number of arguments.)
\begin{lemma}[Performance of tree ensemble channel on BEC]
Consider a degree distribution pair $(\lambda,\rho)$ and a real number
$0 < \alpha < 1$. 
Define $x_0 = \alpha$ and for $\ell \ge
1$, 
\begin{equation}
\label{eq:bec-irreg-recursion}
x_{\ell} = \alpha \lambda(1 - \rho(1-x_{\ell-1}))  \ . 
\end{equation}
Then, for the BEC with erasure probability $\alpha$, for every $\ell \ge 1$, we
  have $P_{\cal T}^{(\lambda,\rho)}(\ell) = x_{\ell}$.
\end{lemma}
\begin{proof}
The proof follows along the lines of the recursion
(\ref{eq:bec-recursion}) that we established for the regular case.
The case $\ell=0$ is clear since the initial variable-to-check message
equals the received value which equals an erasure with probability
$\alpha$. Assume that for $0 \le i < \ell$, $P_{\cal
  T}^{(\lambda,\rho)}(i) = x_i$. In the $\ell$'th iteration, a
check-to-variable node message sent by a degree $i$ check node is the
erasure message if any of the $(i-1)$ incoming messages is an erasure,
an event that occurs with probability $1-(1-x_{\ell-1})^{i-1}$ (since
the incoming messages are independent and each is an erasure with
probability $x_{\ell-1}$ by induction). Since the edge has probability
$\rho_i$ to be connected to a check node of degree $i$, the erasure
probability of a check-to-variable message in the $\ell$'th iteration
for a randomly chosen edge is equal to $\sum_i \rho_i (1 -
(1-x_{\ell-1})^{i-1}) = 1 - \rho(1-x_{\ell-1})$. Now consider a
variable-to-check message in the $\ell$'th iteration sent by a
variable node of degree $i$. This is an erasure iff the node was
originally erased and each of the $(i-1)$ incoming messages are
erasures. Thus it is an erasure with probability $\alpha
(1-\rho(1-x_{\ell-1}))^{i-1}$. Averaging over the edge degree
distribution $\lambda(\cdot)$, we have $P_{\cal
  T}^{(\lambda,\rho)}(\ell) = \alpha \lambda(1-\rho(1-x_{\ell-1})) =
x_{\ell}$.
\end{proof}

The following lemma yields the threshold erasure probability for a
given degree distribution pair $(\lambda,\rho)$. The proof is
identical to Lemma~\ref{lem:bec-reg-threshold} --- we simply use the
recursion (\ref{eq:bec-irreg-recursion}) in place of
(\ref{eq:bec-recursion}). Note that Lemma~\ref{lem:bec-reg-threshold}
is a special case when $\lambda(z) = z^{d_v-1}$ and $\rho(z) =
z^{d_c-1}$. 

\begin{lemma}
\label{lem:bec-irreg-threshold}
For the BEC, the threshold erasure probability
$\a^{\bp}(\lambda,\rho)$ below which the above iterative message
passing algorithm leads to vanishing bit-erasure probability as the
number of iterations grows is given by
\begin{equation}
\label{eq:bec-irreg-threshold}
\a^{\bp}(\lambda,\rho) = \min_{x \in [0,1]}
\frac{x}{\lambda(1-\rho(1-x))} \ . 
\end{equation}
\end{lemma}

\subsection{Capacity achieving distributions for the BEC}

Having analyzed the performance possible on the BEC for a given degree
distribution pair $(\lambda,\rho)$, we now turn to the question of
what pairs $(\lambda,\rho)$, if any, have a threshold approaching
capacity. Recalling the designed rate from (\ref{eq:irreg-rate}), the
goal is to find $(\lambda,\rho)$ for which $\a^{\bp}(\lambda,\rho)
\approx \frac{\int_0^1 \rho(z) dz}{\int_0^1
  \lambda(z) dz}$.

We now discuss a recipe for constructing such degree
distributions, as discussed in \cite{OS02} and
\cite[Sec. 2.9.11]{RU-book} (we follow the latter description closely).
In the following we use parameters $\theta > 0$ and a positive
integer $N$ that will be fixed later.
Let ${\cal D}$ be the space of non-zero functions $h : [0,1)
  \rightarrow \R^+$ which are analytic around zero with a Taylor
  series expansion comprising of non-negative coefficients.
Pick functions $\hat{\lambda}_\theta(x) \in {\cal D}$ and $\rho_\theta(x) \in
{\cal D}$ that satisfy $\rho_\theta(1) = 1$ and 
\begin{equation}
\label{eq:rho-cond}
\hat{\lambda}_\theta(1 - \rho_\theta(1-x)) = x \ ,\quad \forall x
\in [0,1) \ . 
\end{equation}
Here are two example choices of such functions:
\begin{enumerate}
\item Heavy-Tail Poisson Distribution~\cite{LMSS}, dubbed ``Tornado
 sequence'' in the literature. Here we take
\begin{eqnarray*}
 \hat{\lambda}_\theta(x) & = &
  \frac{- \ln (1-x)}{\theta} = \frac{1}{\theta} \sum_{i=1}^{\infty}
  \frac{x^i}{i} \ , \mbox{ and} \\
\rho_\theta(x) & = & e^{\theta(x-1)} = e^{-\theta} \sum_{i=0}^\infty
  \frac{\theta^i x^i}{i!} \ . 
\end{eqnarray*}
\item Check-concentrated degree distribution~\cite{shok-bec}. Here for
  $\theta \in (0,1)$ so that $1/\theta$ is an integer, we take
\begin{eqnarray*}
 \hat{\lambda}_\theta(x) & = & 1- (1-x)^\theta = \sum_{i=1}^\infty
  {{\theta} \choose i} (-1)^{i-1} x^i \ , \mbox{ and} \\
\rho_\theta(x) & = & x^{1/\theta} \ .
\end{eqnarray*}
\end{enumerate}

Let $\hat{\lambda}^{(N)}_\theta(x)$ be the function consisting of the first
$N$ terms (up to the $x^{N-1}$ term) of the Taylor series expansion of
$\hat{\lambda}_\theta(x)$ around zero, and define the normalized function
$\lambda_\theta^{(N)}(x) =
\frac{\hat{\lambda}^{(N)}_\theta(x)}{\hat{\lambda}^{(N)}_\theta(1)}$ 
(for large enough $N$, $\hat{\lambda}^{(N)}_\theta(1) > 0$, and so this
polynomial has positive coefficients).
For suitable parameters $N,\theta$, the pair
$(\lambda_\theta^{(N)},\rho_\theta)$ will be our candidate degree
distribution pair.\footnote{If the power series expansion of
  $\rho_\theta(x)$ is infinite, one can truncate it at a sufficiently high
  term and claimed bound on threshold still applies. Of course for the
  check-concentrated distribution, this is not an issue!}
The non-negativity of the Taylor series coefficients of
$\hat{\lambda}_\theta(x)$ implies that for $x \in [0,1]$, 
$\hat{\lambda}_\theta(x) \ge \lambda^{(N)}_\theta(x)$, which together
with (\ref{eq:rho-cond}) gives
\[ x = \hat{\lambda}_\theta(1 - \rho_\theta(1-x)) \ge
\hat{\lambda}^{(N)}_\theta(1 - \rho_\theta(1-x)) =
\hat{\lambda}^{(N)}_\theta(1) 
 \lambda^{(N)}_\theta(1 - \rho_\theta(1-x)) \ . \]
By the characterization of the threshold in
Lemma~\ref{lem:bec-irreg-threshold}, it follows that $\a^{\bp}(\lambda^{(N)}_\theta,\rho_\theta) \ge
\hat{\lambda}^{(N)}_\theta(1)$.  Note that the designed rate equals
\[ r = r(\lambda^{(N)}_\theta,\rho_\theta) =  1 - \frac{\int_0^1 \rho_\theta(z) dz}{\int_0^1
  \lambda^{(N)}_\theta(z) dz} = 
1  - \hat{\lambda}^{(N)}_\theta(1)
 \frac{\int_0^1 \rho_\theta(z) dz}{\int_0^1
  \hat{\lambda}^{(N)}_\theta(z) dz} \ .  \]

Therefore, given a target erasure probability $\a$, to communicate at
rates close to capacity $1-\a$, the functions
$\hat{\lambda}^{(N)}_\theta$ and $\rho_\theta$ must satisfy 
\begin{equation}
\hat{\lambda}^{(N)}_\theta(1) \approx \a \quad \mbox{ and } 
 \frac{\int_0^1 \rho_\theta(z) dz}{\int_0^1
  \hat{\lambda}^{(N)}_\theta(z) dz} \to 1 \mbox{ as $N \to \infty$} \
  . 
\end{equation}

For example, for the Tornado sequence, $\hat{\lambda}^{(N)}_\theta(1)
= \frac{1}{\theta} \sum_{i=1}^{N-1} \frac{1}{i} =
\frac{\mathsf{H}(N-1)}{\theta}$ where $\mathsf{H}(m)$ is the Harmonic
function. Hence, picking $\theta = \frac{\mathsf{H}(N-1)}{\a}$ ensures
that the threshold is at least $\a$. We have $\int_{0}^{1}
\hat{\lambda}^{(N)}_\theta(z) dz = \frac{1}{\theta} \sum_{i=1}^{N-1}
\frac{1}{i(i+1)} = \frac{N-1}{\theta N}$, and $\int_0^1 \rho_\theta(z)
dz = \frac{1 - e^{-\theta}}{\theta}$.  Therefore, $ \frac{\int_0^1
\rho_\theta(z) dz}{\int_0^1 \hat{\lambda}^{(N)}_\theta(z) dz} =
(1-e^{-\mathsf{H}(N-1)/\a}) (1-1/N) \to 1$ as $N \to \infty$, as
desired. Thus the degree distribution pair is explicitly given by
\[ \lambda^{(N)}(x) = \frac{1}{\mathsf{H}(N-1)} \sum_{i=1}^{N-1}
\frac{x^i}{i} \ , \quad \rho^{(N)}(x) = e^{\frac{\mathsf{H}(N-1)}{\a}
  (x-1)} \ . \]

Note that picking $N \approx 1/\eps$ yields a rate $(1-\eps)\a$ for
reliable communication on $\bec_\a$. The average variable node degree
equals $\frac{1}{\int_0^1 \lambda^{(N)}(z) dz} \approx \mathsf{H}(N-1)
\approx \ln N$. Therefore, we conclude that we achieve a rate within a
multiplicative factor $(1-\eps)$ of capacity with decoding complexity
$O(n \log (1/\eps))$.

For the check-concentrated distribution, if we want to achieve
$\a^{\bp}(\lambda^{(N)}_\theta,\rho_\theta) \ge \a$ and a rate $r \ge
(1-\eps) \a$, then it turns out that the choice $N \approx 1/\eps$ and
$1/\theta = \lceil \frac{\ln N}{- \ln (1-\a)} \rceil$ works. In
particular, this means that the factor graph has at most $O(n
\log(1/\eps))$ edges, and hence the ``Peeling decoder'' will again run in
$O(n \log(1/\eps))$ time.  

One might wonder that among the various capacity achieving degree
distributions that might exist for the BEC, which one is the ``best''
choice?  It turns out that in order to achieve a fraction $(1-\eps)$
of capacity, the average degree of the factor graph has to be
$\Omega(\ln (1/\eps))$. This is shown in \cite{SU-checkoptimal} using
a variant of Gallager's argument for lower bounding the gap to
capacity of LDPC codes. In fact, rather precise lower bounds on the
sparsity of the factor graph are known, and the check-concentrated
distribution is optimal in the sense that it matches these
bounds very closely; see \cite{SU-checkoptimal} for the detailed
calculations.

In light of the above, it might seem that check-concentrated
distributions are the final word in terms of the
performance-complexity trade-off. While this is true in this framework of
decoding LDPC codes, it turns out by using more complicated graph
based codes, called Irregular Repeat-Accumulate Codes, even better
trade-offs are possible~\cite{PSU-capbounded}. We will briefly return
to this aspect in Section~\ref{sec:ira}.

\subsection{Extensions to channels with errors}

Spurred by the remarkable success of \cite{LMSS} in achieving capacity
of the BEC, Luby {\it et al}~\cite{LMSS-errors} investigated the
performance of 
irregular LDPC codes for the BSC.

In particular, they considered the natural extension of Gallager's
Algorithm B to irregular graphs, where in iteration $i$, a variable
node of degree $j$ uses a threshold $b_{i,j}$ for flipping its value.
Applying essentially the same arguments as in
Section~\ref{sec:gallager-algoB}, but accounting for the degree
distributions, one gets the following recurrence for the expected
fraction $p_\ell$ of incorrect variable-to-check messages in the
$\ell$'th iteration:
\begin{align*}
{p}_{i+1} & =  {p}_0 - {p}_0 
\sum_{j=1}^{d_v^{\max}} \sum_{t=b_{i+1,j}}^{j}
      {\textstyle {{j-1} \choose t}}
      \left(\frac{1+\rho(1-2{p}_i)}{2}\right)^t
\left(\frac{1-\rho(1-2{p}_i)}{2}\right)^{j-1-t}  
\\ 
& +
	  (1-{p}_0)  \sum_{j=1}^{d_v^{\max}} \sum_{t=b_{i+1,j}}^{j}
      {\textstyle {{j-1} \choose t}}
      \left(\frac{1+\rho(1-2{p}_i)}{2}\right)^{j-1-t}
\left(\frac{1-\rho(1-2{p}_i)}{2}\right)^{t}
\end{align*}
As with the regular case, the cut-off value $b_{i+1,j}$ can then be chosen to minimize the value
of $p_{i+1}$, which is given by the smallest integer
for which
\[ \frac{1-{p}_0}{{p}_0} \le
\left(
\frac{1+\rho(1-2{p}_i)}{1-\rho(1-2{p}_i)}\right)^{2b_{i+1,j}-j+1}
\ . \] 
Note that $2b_{i+1,j}-j+1 = b_{i+1,j} - (j-1 -b_{i+1,j})$ equals the
difference between the number of check nodes that agree in the majority
and the number that agree in the minority. Therefore, a variable
node's decision in each iteration depends on whether this difference
is above a certain threshold, regardless of its degree.

Based on this, the authors of \cite{LMSS-errors} develop a linear
programming approach to find a good $\lambda$ given a distribution
$\rho$, and use this to construct some good degree distributions. Then
using the above recurrence they estimate the theoretically achievable
threshold crossover probability. Following the development of the
density evolution algorithm to track the performance of belief
propagation decoding~\cite{RU01}, the authors of \cite{RSU01} used
optimization techniques to find good irregular degree distributions
for belief propagation decoding. The BIAWGN channel was the primary
focus in \cite{RSU01}, but the authors also list a few examples
that demonstrate the promise of the techniques for other channels. In
particular, for the BSC with rate $1/2$, they report a degree
distribution pair with maximum variable node degree $75$ and
check-node distribution $\rho(x) = 0.25 x^9 + 0.75 x^{10}$ for which
the computed threshold is $0.106$, which is quite close to the Shannon
capacity limit $0.11$. The techniques were further refined and codes
with rate $1/2$ and a threshold of $\sigma^* \approx 0.9781$ (whose
SNR is within $0.0045$ dB of capacity) were reported for the BIAWGN in
\cite{CFRU01} --- these codes use only two different check node degrees
$j,j+1$ for some integer $j \ge 2$.

\section{Linear encoding time and Repeat-Accumulate Codes}
\label{sec:ira}
The linear decoding complexity of LDPC codes is one of their
attractive features. Being linear codes, they generically admit
quadratic time encoding. In this section, we briefly discuss how the
encoding complexity can be improved, and give pointers to where
results in this vein can be found in more detail.

The original Tornado codes paper~\cite{LMSS} achieved linear time
encoding using a cascade of several low-density generator matrix
(LDGM) codes. In LDGM codes, the ``factor'' graph is actually used to
compute actual check bits from the $k$ message bits (instead of
specifying parity checks that the codeword bits must obey). Due to the
sparse nature of the graph, the check bits can be computed in linear
time. These check bits are then used as message bits for the next
layer, and so on, till the number of check bits becomes
$O(\sqrt{k})$. These final set of check bits are encoded using a
quadratic time encodable linear code.

\smallskip 
We now mention an alternate approach to achieve linear time encoding
for LDPC codes themselves (and not a cascaded variant as in
\cite{LMSS}), based on finding a sparse parity check matrix with
additional nice properties.  Let $H \in \F_2^{m \times n}$ be the
parity check matrix of an LDPC code of dimension $n-m$. By means of
row and column operations, we can convert $H$ into a form $\tilde{H}$
where the last $m$ columns are linearly independent, and moreover the
$m \times m$ submatrix consisting of the last $m$ columns is lower
triangular (with $1$'s on the diagonal). Using $\tilde{H}$, it is a
simple matter of ``back-substitution'' to compute the $m$ parity bits
corresponding to the $n-m$ information bits (the encoding is {\em
systematic}). The complexity of this encoding is governed by the
number of $1$'s in $\tilde{H}$. In general, however, when we begin
with a sparse $H$, the resulting matrix $\tilde{H}$ is no longer
sparse. In a beautiful paper~\cite{RU-linencoding}, Richardson and
Urbanke propose finding an ``approximate'' lower triangulation of the
parity check matrix that is still sparse. The idea is to make the top
right $(m-g) \times (m-g)$ corner of the matrix lower triangular for
some small ``gap'' parameter $g$. The encoding can be done in
$O(n+g^2)$ time, which is linear if $g = O(\sqrt{n})$. Remarkably, for
several distribution pairs $(\lambda,\rho)$, including all the
optimized ones listed in \cite{RSU01}, it is shown in
\cite{RU-linencoding} that, with high probability over the choice of
the code from the ensemble $\ldpc(n,\lambda,\rho)$, a gap of
$O(\sqrt{n})$ can in fact be achieved, thus leading to linear encoding
complexity!

\medskip 
Yet another approach to achieve linear encoding complexity that we
would like to focus on (as it has some additional applications), is to
use Irregular Repeat-Accumulate (IRA) codes. IRA codes were introduced
by Jin, Khandekar and McEliece in \cite{JKM-IRA}, by generalizing the
notion of Repeat-Accumulate codes from \cite{DJM-allerton} in
conjunction with ideas from the study of irregular LDPC codes.

IRA codes are defined as follows. Let $(\lambda,\rho)$ be a degree
distribution pair. Pick a random bipartite graph $G$ with $k$ {\em
information} nodes on left (with a fraction $\lambda_i$ of the edges
being incident on information nodes of degree $i$), and $n > k$ {\em
check} nodes on the right (with a fraction $\rho_i$ of the edges
incident being incident on check nodes of degree $i$). Actually, it
turns out that one can pick the graph to be regular on the check node
side and still achieve capacity, so we can even restrict ourselves to
check-degree distributions given by $\rho_a=1$ for some integer
$a$. Using $G$, the encoding of the IRA code (of dimension $k$ and
block length $n$) proceeds as follows:
\begin{itemize}
\item Place the $k$ message bits on the $k$ information nodes.
\item For $1 \le i \le n$, at the $i$'th check node, compute the bit
  $v_i \in \{1,-1\}$ which equals the parity (i.e., product, in $\pm 1$
  notation) of the message bits placed on its
  neighbors. 
\item (Accumulation step) Output the codeword $(w_1,w_2,\dots,w_n)$
  where $w_j = \prod_{i=1}^j v_i$. In other words, we accumulate the
  parities of the prefixes of the bit sequence $(v_1,v_2,\dots,v_n)$.
\end{itemize}

Note that the encoding takes $O(n)$ time. Each of the check nodes
has constant degree, and thus the $v_i$'s can be computed in linear
time. The accumulation step can then be performed using additional
$O(n)$ operations.

It is not hard to show that the rate of the IRA code corresponding to
 a pair $(\lambda,\rho)$ as defined above equals 
$\frac{\int_{0}^1
 \lambda(z) dz}{\int_0^1 \rho(z) dz}$.

A natural iterative decoding algorithm for IRA codes is presented and
analyzed in \cite{DJM-allerton} (a description also appears in
\cite{PSU-capbounded}). The iterative algorithm uses a graphical model
for message passing that includes the above bipartite graph $G$
connecting information nodes to check nodes, juxtaposed with another
bipartite graph connecting the check nodes to $n$ {\em code} nodes
labeled $x_1,x_2,\dots,x_n$. In this graph, which is intended to
reflect the accumulation process, 
code node $x_i$ for $1 \le i < n$ is connected to the $i$'th and
$(i+1)$'th check nodes (ones where $v_i,v_{i+1}$ are computed), and
node $x_n$ is connected to the check node where $v_n$ is computed.

It is proved (see \cite[Sec. 2]{PSU-capbounded}) that for the above
{\em non-systematic} IRA codes, the iterative decoding on $\bec_\alpha$ converges to vanishing
bit-erasure probability as the block length $n \to \infty$, provided
\begin{equation}
\label{eq:ira-cond}
\lambda \left( 1 - \left[ \frac{1-\a}{1- \a R(1-x)} \right]^2
\rho(1-x) \right) < x \quad \forall x \in (0,1] \ . 
\end{equation}
In the above $R(x) = \sum_{i=1}^\infty R_i x^i$ is the power series
whose coefficient $R_i$ equals the fraction of check nodes that are
connected to $i$ information nodes in $G$. Recalling
(\ref{eq:inverse-rel}), we have $R(x) =
 \frac{\int_{0}^x \rho(z)
dz}{\int_0^1 \rho(z) dz}$. 

Using the above characterization, degree distribution pairs
$(\lambda,\rho)$ for IRA codes that achieve the capacity of the BEC
have been found in \cite{DJM-allerton,SaU04}.\footnote{Actually, 
  these papers work with a {\em systematic} version of IRA where
  the codeword includes the message bits in addition to the
  accumulated check bits $x_1,\dots,x_n$. Such systematic codes have rate equal
  to $\left( 1 + \frac{\int_{0}^1
\rho(z) dz}{\int_0^1 \lambda(z) dz}\right)^{-1}$, and the decoding
  success condition (\ref{eq:ira-cond}) for them is slightly different, with a
  factor $\a$ multiplying the $\lambda(\cdot)$ term on the left hand
  side.}
In particular, we want to draw attention to the construction in
\cite{PSU-capbounded} with $\rho(x) = x^2$ that can achieve a rate of
$(1-\eps)(1-\a)$, i.e., within a $(1-\eps)$ multiplicative factor of
the capacity of the BEC, for $\a \in [0,0.95]$.\footnote{The claim is
  conjectured to hold also for $\a \in (0.95,1)$.}
 Since $\rho(x) = x^2$, all check nodes are connected to exactly $3$
information nodes. Together with the two code nodes they are connected
to, each check node has degree $5$ in the graphical model used for
iterative decoding. The total number of edges in graphical model is
thus $5n$, and this means that the complexity of the encoder as well
as the ``Peeling'' implementation of the decoder is at most $5n$. In
other words, the complexity per codeword bit of encoding and decoding
is bounded by an absolute constant, independent of the gap $\eps$ to
capacity.

\section{Summary}

We have seen that LDPC codes together with natural message-passing
algorithms constitute a powerful approach for the channel coding
problem and to approach the capacity of a variety of channels. For the
particularly simple binary erasure channel, irregular LDPC codes with
carefully tailored degree distributions can be used to communicate at
rates arbitrarily close to Shannon capacity. Despite the impressive
strides in the asymptotic analysis of iterative decoding of irregular
LDPC codes, for all nontrivial channels except for the BEC, it is
still unknown if there exist sequences of degree distributions that
can get arbitrarily close to the Shannon limit.  By optimizing degree
distributions numerically and then computing their threshold (either
using explicit recurrences or using the density evolution algorithm),
various rather excellent bounds on thresholds are known for the BSC
and BIAWGN. These, however, still do not come close to answering the
big theoretical open question on whether there are capacity-achieving
ensembles of irregular LDPC codes (say for the BSC), nor do they
provide much insight into their structure. 

For irregular LDPC codes, we have explicit sequences of {\em
  ensembles} of codes that achieve the capacity of the BEC (and come
  pretty close for the BSC and the BIAWGN channel). The codes
  themselves are not fully explicit, but rather sampled from the
  ensemble. While the concentration bounds guarantee that almost all
  codes from the ensemble are likely to be good, it may still be nice
  to have an explicit family of codes (rather than ensembles) with
  these properties. Even for achieving capacity of the BEC, the only
  known ``explicit'' codes require a brute-force search for a rather
  large constant sized code, and the dependence of the decoding
  complexity on the gap $\eps$ to capacity is not as good as for
  irregular LDPC ensembles. For the case of errors, achieving a
  polynomial dependence on the gap $\eps$ to capacity remains an
  important challenge.

\end{document}